# Bizard: A Community-Driven Platform for Accelerating and Enhancing Biomedical Data Visualization

Kexin Li[1]†, Hu Zheng[2]†, Kexin Huang[3]†, Yinying Chai[4]†, Yujie Peng[4]†, Chunyang Wang[5]†, Xuyang Yi[6]†, Zilun Jin[7]†, Hong Yang[1]†, Yun Peng[8]†, Ying Shi[4]†, Xinhe Lu[9], Jiarui Bian[3], Yirun Wang[3], Rongao Kou[10], Demin Gao[11], Hanbo Zhao[12], Juan Zhang[13], Dan Huang[14], Kaiyu Zhu[15], Chenxu Wu[16], Zeruo Yang[17], Zheng Kuang[18], Mo Liu[19], Zhiwei Bao[20], Yuzhong Peng[21], Benben Miao[22], Jianming Zeng[21]*, Jianfeng Li[23,24]*, Peng Luo[1]*, Shixiang Wang[5]*

[1]Department of Oncology, Zhujiang Hospital, The Second School of Clinical Medicine, Southern Medical University, Guangzhou, China

[2]College of Veterinary Medicine, Huazhong Agricultural University, Wuhan, Hubei, China

[3]The First School of Clinical Medicine, Southern Medical University, Guangzhou, Guangdong, China

[4]The Second School of Clinical Medicine, Southern Medical University, Guangzhou, Guangdong, China

[5]Department of Biomedical Informatics, School of Life Sciences, Central South University, Changsha, China

[6]Department of Thoracic Surgery, The Second Xiangya Hospital of Central South University, Changsha, China

[7]State Key Laboratory of Crop Stress Adaptation and Improvement, School of Life Sciences, Henan University, Kaifeng, Henan, China

[8]Peking University People's hospital, Peking University, Beijing, China

[9]Key Laboratory of Environment and Health, Ministry of Education & Ministry of Environmental Protection, and State Key Laboratory of Environmental Health, School of Public Health, Tongji Medical College, Huazhong University of Science and Technology, Wuhan, Hubei, China.

[10]School of Life Sciences, Westlake University, Hangzhou, Zhejiang, China

[11]Centre for Chinese Medicine Drug Development Limited, Hong Kong Baptist University, Hong Kong SAR, China

[12]Kunpeng Institute of Modern Agriculture, Foshan, Guangdong, China

[13]Yiyan Biopharmaceutical Technology Co., Ltd., Zhongshan, Guangdong, China

[14]College of Natural Sciences, Pusan National University, Busan, South Korea

[15]Center for Gastrointestinal Surgery, The First Affiliated Hospital, Sun Yat-Sen University, Guangzhou, China

[16]School of Life Science and Technology, ShanghaiTech University, Shanghai, China

[17]Center for Quantitative Biology, Peking University, Beijing, China

[18]Department of Immunology, Peking University Health Science Center, Peking University, Center for Human Disease Genomics, No. 38, Xueyuan Road, Beijing, China

[19]Sino-French Hoffmann Institute, School of Basic Medical Sciences, Guangzhou Medical University, Guangzhou, Guangdong, China

[20]College of Biomedical Engineering and Instrument Science, Ministry of Education Key Laboratory of Biomedical Engineering, Zhejiang University, Hangzhou, China

[21]Faculty of Health Sciences, University of Macau, Avenida de Universidade, Taipa, Macau SAR, China

[22]State Key Laboratory of Mariculture Breeding, College of Ocean and Earth Sciences, Xiamen University, Xiamen, China

[23]Shanghai Institute of Hematology, State Key Laboratory of Medical Genomics, National Research Center for Translational Medicine at Shanghai, Ruijin Hospital Affiliated to Shanghai Jiao Tong University School of Medicine, Shanghai, China

[24]School of Life Sciences and Biotechnology, Shanghai Jiao Tong University, Shanghai, China

†These authors contributed equally to this work.

*Correspondence:

Jianming Zeng, yb77613@umac.mo;

Jianfeng Li, lee_jianfeng@sjtu.edu.cn;

Peng Luo, luopeng@smu.edu.cn;

Shixiang Wang, wangshx@csu.edu.cn.

## Abstract

Biomedical research increasingly relies on heterogeneous, high-dimensional datasets, yet effective visualization remains hindered by fragmented code resources, steep programming barriers, and limited domain-specific guidance. Bizard is an open-source visualization code repository engineered to streamline data analysis in biomedical research. It aggregates a diverse array of executable visualization scripts, empowering researchers to select and tailor optimal graphical methods for their specific investigative demands. The platform features an intuitive interface equipped with sophisticated browsing and filtering capabilities, exhaustive tutorials, and interactive discussion forums that foster knowledge dissemination. Through its community-driven paradigm, Bizard promotes continual refinement and functional expansion, establishing itself as an essential resource for elevating biomedical data visualization and analytical standards. By harnessing Bizard's infrastructure, researchers can augment their visualization proficiency, propel methodological progress, and enhance interpretive rigor, ultimately accelerating precision medicine and personalized therapeutics. Bizard is freely accessible at https://openbiox.github.io/Bizard/.

## Summary

- An open source professional visualization platform for the biomedical research.
- Visualization solutions for approximately 750 graphics, including basic, omics, and clinical analysis.
- User-friendly website design and bilingual support.
- Community-driven operation mechanism and automated deployment based on GitHub.

## 1 Introduction

In the era of big biomedical data, visualization serves as an indispensable analytical lens, transforming opaque, high-dimensional datasets into intuitive, hypothesis-generating narratives that propel both fundamental discovery and clinical decision-making [1-3]. The rapid advancement of high-throughput sequencing technologies, high-resolution imaging techniques, and various omics technologies has generated massive and complex biomedical data. These datasets contain immense potential for unraveling the mysteries of life, understanding disease mechanisms, and developing diagnostic and therapeutic approaches [4, 5]. However, the effective extraction of meaningful information from these intricate datasets and their transformation into clear, accurate, and compelling visual representations has become a significant challenge confronting every biomedical researcher [6, 7].

The R ecosystem, with its unparalleled statistical prowess, expansive package universe, and graphic flexibility, has emerged as the preeminent environment for such tasks [8-10]. However, achieving high-quality visualizations is no easy task. Researchers often face multiple challenges: First, the technical barrier is high. Many powerful visualization tools (such as ggplot2) require a solid programming foundation, posing a significant obstacle for biology or medical researchers lacking computational backgrounds [11]. Second, resources are scattered and disorganized. Although the internet hosts a vast amount of fragmented visualization codes and tutorials, their quality varies greatly, and finding, filtering, and adapting these resources demand considerable time and effort from researchers. Additionally, personalized customization is difficult. Even when basic code is found, adjusting and optimizing it according to specific biomedical data types and specific publication requirements often requires in-depth understanding and repeated debugging. Finally, there is a lack of deep integration with domain knowledge. General visualization tutorials often fail to provide interpretations and application cases within a biomedical context, making it difficult for researchers to closely connect the charts with actual scientific questions [12, 13].

Currently, while several excellent visualization tools are available for analyzing diverse data in biomedicine, the rapid emergence of numerous plotting packages and their associated code implementations often poses significant challenges for researchers [14-21]. They must select the most suitable visualization methods based on their specific needs and implement corresponding customizations. This process demands advanced programming skills, continuous interactive exploration, iterative experimentation, optimization, and effective cross-disciplinary communication. Although many online visualization platforms have reduced the requirement for programming skills, these tools often lack transparent code and are difficult to modify for personalized needs [22-27]. The R Graph Gallery has made important contributions to R language data visualization, but it primarily focuses on providing diverse chart types or basic plotting functions, lacking biomedical-specific visualizations (**Table 1**). FigureYa has significantly contributed to data visualization in the biomedical field [28], yet its website design and community operations still require improvement. A comprehensive, ecosystem-level platform that integrates code storage, tutorial learning, data adaptation, and community exchange remains scarce. Researchers need an environment that not only enables "ready-to-use" solutions but also supports "learning while using," "understanding through application," and "collaborative creation." Such a platform would systematically address visualization challenges throughout the complete workflow from data to insights, and from insights to publication.

To address this challenge, here we present Bizard (https://openbiox.github.io/Bizard/), a comprehensive visualization code repository based on community collaboration. Bizard serves as an all-encompassing reference data visualization platform for biomedical researchers, offering not only a wide array of code but also reproducible guidance documents. It provides advanced browsing and filtering mechanisms, curated reference codes, examples of diverse visualization methods, and interactive discussion forums to facilitate peer knowledge exchange. Ultimately, Bizard aims to enhance data-driven research and cultivate a continuously growing community.

# 2 Methods

## 2.1 Methods for Bizard website construction

### 2.1.1 Website front-end layout design

The Bizard website's front-end architecture is constructed using Quarto (https://github.com/quarto-dev/quarto-cli), an open-source publishing system that offers seamless compatibility with R computational environments. This technical foundation enables the direct integration of executable code chunks within the documentation, allowing for dynamic generation of visualization examples and ensuring consistency between code implementation and graphical output. The layout design implements a responsive multi-column structure that adapts intelligently to various viewing devices. Based on CSS Grid layout technology, the system automatically adjusts content arrangement according to screen width. The navigation system features a hierarchical architecture with a collapsible navigation bar on the left and a sidebar on the right. The left navigation incorporates breadcrumb trails that dynamically update based on user location, while the right sidebar allows users to navigate through the sections of a graphic tutorial.

### 2.1.2 Graph Gallery page design

The Graph Gallery feature was introduced to comprehensively address challenges arising from the rapidly growing number of graphics, including difficulties in locating specific graphics, management disorganization, and inefficient usage. By leveraging the advanced table management tool reactable (https://github.com/glin/reactable), Bizard has integrated all graphics and rendered the table using JavaScript and CSS front-end code. In this tabular interface, each graphic is presented as a clear entry with key attributes: Graphic ID, Name, Description, and Type. Users can accurately locate the content they need from the extensive graphics collection through filtering, sorting, and keyword search, significantly enhancing the efficiency of information retrieval.

### 2.1.3 Bilingual implementation

To better serve our global user base and eliminate barriers caused by language differences, Bizard has implemented seamless switching between Chinese and English for all documentation on the platform. We achieve this using babelquarto (https://github.com/ropensci-review-tools/babelquarto/), a multilingual rendering tool fully compatible with the Quarto framework. For each graphic tutorial, we provide both Chinese and English documents that have been reviewed and proofread by Bizard operations team, ensuring translation accuracy. This initiative aims to create a truly international and inclusive website platform. This means users can freely switch between English and Chinese documentation with the click of a button, easily accessing the information they need.

## 2.2 Methods for Bizard graphic documents organization

### 2.2.1 Graphics collection

Bizard integrated visualization code from a variety of sources, including traditional graphic codes from the R Graph Gallery (https://r-graph-gallery.com/) and innovative visualization solutions from experts in related fields [29-33]. To enrich the content of Bizard's graphic tutorials and leverage the advantages of its community-driven operational model, Bizard consistently solicits biomedical research-related graphics from community members worldwide. During Bizard's first round of chart collection, 31 graphic types were reviewed and evaluated by the project team for inclusion in the Bizard platform [34-38]. Notably, Bizard has also integrated 177 graphics from the Hiplot platform [24], an interactive biomedical data visualization application developed by the Openbiox community. Since 2019, the Hiplot platform has implemented hundreds of biomedical visualization plugins covering areas such as basic statistics, multi-omics analysis, regression analysis, cluster analysis, dimensionality reduction analysis, meta-analysis, survival analysis, and risk model construction. The inclusion of these graphics not only enriches Bizard's project content but also provides source code tutorial analysis for Hiplot's visualization plugins, helping users better understand and utilize these tools.

### 2.2.2 Graphics classification

To address the classification confusion caused by the sheer number and overlapping dimensions of the hundreds of graphics on the Bizard platform, Bizard has implemented a graphic directory management system based on a three-level classification structure. The first level comprises major graphic categories such as Basics, Omics, Clinics, and Hiplot. The second level organizes graphics by functional types, including Distribution, Correlation, Ranking, and Composition, etc. The third level consists of individual graphic types, such as Violin Plot, Density Plot, Histogram, and Box Plot, etc. The significant advantage of this hierarchical classification structure is that it groups graphics according to their functions and categories, transforming scattered graphics into an interconnected network. Ultimately, this approach constructs a comprehensive graphic visualization knowledge base.

#### 2.2.3 Graphics data organization

In data analysis and visualization work, efficiently and systematically organizing and hosting example datasets is a crucial step. To ensure that the data can be accessed remotely and securely, Bizard organizes example datasets according to specific standards and processes. Data is organized into easily readable data frame or matrix formats, with each column containing headers, and missing values are removed. To facilitate user accessibility, we removed irrelevant, duplicate, and blank rows and columns to keep the sample data files as small as possible. After data organization, these datasets were uploaded to Tencent Cloud and accessed via uniform resource locators (URLs). Bizard provides example datasets URLs in each tutorial, which can be directly loaded through R code.

#### 2.2.4 Standardization of graphic documents

Bizard project has established unified and standardized code and documentation format. This not only facilitates collaborative development among community contributors but also supports biomedical researchers in learning and utilization. A standard Bizard document should include the following sections: (1) Example: This section provides basic usage examples for each graphic along with descriptive information; (2) Setup: This section explains how to setup the programming environment required for plotting; (3) Data Preparation: This section introduces the data used for visualization; (4) Visualization: This section demonstrates multiple approaches to creating the graphics; (5) Application: This section presents various application scenarios for the visualizations; (6) Reference: This section provides relevant literature and documentation sources.

## 2.3 Methods for Bizard community management

### 2.3.1 Comment system construction

To build an efficient and scalable community communication platform, Bizard fully leverages the advantages of the GitHub ecosystem by utilizing the giscus (https://github.com/giscus/giscus) tool to achieve seamless integration between the comment system and code repository. The giscus system is developed based on the GitHub Discussion API and built using a modern web technology stack. The front-end employs the React framework to create a dynamic user interface, while GitHub OAuth is used for user authentication, ensuring the security and reliability of the comment system. The system architecture adopts a client-direct communication model with the GitHub API, avoiding performance bottlenecks associated with intermediate servers while ensuring real-time data synchronization. Each tutorial page is linked to the corresponding GitHub Discussion topic through a unique discussion category ID, guaranteeing that users can seamlessly use the comment function to communicate whenever they encounter problems in any graphic tutorials.

### 2.3.2 Automated build and deployment

To enhance the efficiency of community collaborative project development and eliminate the tediousness and risks of human error associated with manual updates, Bizard has established an efficient and reliable automated build and deployment system using GitHub Actions. When community members push code to the main branch or initiate pull requests, the workflow is automatically triggered. The system first performs code quality checks to ensure compliance with project standards. It then automatically builds the website, configuring the R language environment and necessary system dependencies within the virtual environment provided by GitHub. After administrators review and merge the changes, the merged content is automatically deployed to GitHub Pages. Throughout the automated build and deployment process, GitHub Actions notifies both contributors and administrators of any issues and results via email, providing robust support for rapid iteration and stable operation of the project.

### 2.3.3 Community management based on GitHub

Bizard project fully leverages GitHub's advantages as an open-source collaboration platform to establish a comprehensive community-driven management model. This management model not only promotes the sustainable development of the project but also fosters an active knowledge-sharing ecosystem in biomedical visualization. Bizard adopts a hierarchical community governance structure. Administrators, consisting of project founders and the core team, are responsible for project vision planning and major decision-making. They maintain write access to the code repository and oversee reviewing and merging pull requests. Regular contributors focus on developing and maintaining specific modules, making contributions through pull requests. General users can provide feedback and suggestions through comments during their usage. This pyramid-like structure ensures project quality while lowering the barrier to participation. The Bizard project has also implemented a points-based system to record participants' contributions. The points rules are determined annually through discussions between project initiators and leaders. When submitting articles, authorship order is designated according to the points ranking. Through these open and transparent rules, Bizard builds community trust and recognizes the diverse contributions of its members.

# 3 Results

## 3.1 Overview of Bizard

Bizard is a bioinformatics tool designed to store visualisation code and provide users with a collaborative platform, characterised by its strong openness and comprehensiveness (**Figure 1**). Bizard currently includes approximately 750 graphics across more than 65 chart types (**Supporting Information S1: Table S1**). Leveraging its outstanding data coverage capabilities and broad application scenarios, it provides users diverse options. Focused on the biomedical field, Bizard encompasses common biomedical visualisation codes and provides users with detailed visualisation tutorials and customisation guides. By integrating data from biomedical sources and article reproduction workflows, it facilitates flexible adaptation of visualisations for practical applications. To address user queries and requirements during usage, Bizard has established communication platforms on GitHub and WeChat. Posts on the WeChat platform enable users to conveniently access updates, while the GitHub platform hosts discussion forums where users may freely share opinions and engage in topic-specific exchanges. Additionally, Bizard has integrated DeepWiki and Zread, empowering the platform with AI-powered reading and interactive dialogue capabilities. Users can now utilize natural language chat to intelligently learn and explore all code and tutorial resources within our platform. Bizard warmly welcomes contributions from scholars, anticipating collective efforts to enrich its content structure and deliver superior resources to a broader user base.

### 3.2 Bizard's Workflow

Bizard's workflow comprises four distinct stages: data preparation, fundamental chart generation, biomedical data visualisation, and practical application (**Figure 2**). Bizard first assesses users' requirements for biomedical visualisation, selects appropriate charts, acquires suitable data, and ultimately presents users with tutorial-based explanations of chart functionality, data display (including support for downloading data from public databases), environment configuration, and code interpretation. This enables unrestricted execution of subsequent code. Subsequently, Bizard employs foundational code to demonstrate basic plotting results, including image data and code parameter interpretations. It then provides commonly used parameter adjustments to accommodate personalised requirements derived from user data. Concurrently, Bizard incorporates biomedical datasets to assist users in selecting optimal charts for practical applications. These include common omics data, clinical metrics, and epidemiological information. Each tutorial concludes by demonstrating real-world applications of the chart using authoritative published articles. For select charts, Bizard additionally provides reproducibility tutorials, enabling users to reference professional visualisations for personalised image modifications.

### 3.3 The representative graphics in Bizard

Based on its visual charting capabilities, Bizard's existing charts can be categorised into three types: Basic, Omics, and Clinics, each corresponding to distinct data types and application scenarios (**Figure 3**). The Basic section encompasses common charts such as heatmaps and Venn diagrams, alongside standard analyses like Principal Component Analysis (PCA) and Uniform Manifold Approximation and Projection (UMAP), aiming to provide users with foundational tutorials for diverse scenarios. Simultaneously, diverse parameter adjustments enhance customisation flexibility on these user-friendly foundational plots, enabling users to tailor professional and aesthetically pleasing charts for specific applications. Omics-based plotting encompasses volcano plots and enrichment analyses suited to fundamental multi-omics studies, alongside advanced requirements like motif analysis and chromosome plots. The addition of the Omics module enables users to select more appropriate visualisations for their omics data, while Bizard's customisation features and differential analysis parameters facilitate convenient statistical analysis. The Clinics module incorporates charts such as nomograms and meta forests for analysing clinical data. This module provides tools for analysing clinical data and investigating the association between clinical indicators and disease outcomes. In addition, Bizard has integrated 177 graphics from Hiplot, each linked directly to its corresponding application on Hiplot platform.

### 3.4 The use case of Bizard

Upset plot, a basic graphic on the Bizard website, was selected to illustrate the general operation process of Bizard visualization steps (**Figure 4**). Once entering the Bizard Upset plot page, users can follow these five steps to operate, which include: (1) Example: This section provides an overall introduction of the graphic, describes the content and purpose of graphic, and shows basic usage examples; (2) Setup: This section introduces the system requirements, programming languages, and dependent packages required for ploting. To draw an Upset plot, you need to install and load the UpSetR and ggupset R packages; (3) Data preparation: This section introduces the dataset required for creating the plot. To facilitate users' utilization of the plotting code, Bizard provides example datasets in each tutorial, all of which are uploaded to Tencent Cloud and accessible via online links; (4) Visualization: This section shows the visualization code for plotting. The parameters in Bizard's visualization code have been optimized to help users quickly generate graphics that meet journal publication requirements; (5) Applications and reference: This section introduces the specific applications of graphics in published literature and the references. For example, the Upset plot can be used to display statistics of different diseases in a population and their corresponding survival times.

### 3.5 The community-driven co-operation mechanism of Bizard

Bizard is a GitHub-centric, community-driven platform. Based on GitHub, it is designed to integrate questions discussion, graphics collection, cooperative development, and automatic deployment into a single platform (**Figure 5**).

The Questions Discussion module is built based on GitHub Discussions. GitHub Discussions is a forum-like platform that provides a relaxed and open space for project communication. In Bizard, we built a website comment system based on giscus, which is located at the bottom of the website page. Any user who has questions about the graphic tutorials can start a discussion using the comment box at the bottom of the tutorial page. Once you submit a comment, your question will be automatically announced in the GitHub Discussions section of the Bizard project repository and marked with a link to the graphic tutorial. Any community member who sees your question can contribute to the discussion, and the Bizard operations team will regularly respond to questions in GitHub Discussions.

The Graphics Collection module is built based on GitHub Issues. Bizard is always looking for new graphics from the community. Any user can submit graphics to Bizard by providing the graphic's name, category, example, and related resources via GitHub Issues. Once you submit GitHub Issues, the Bizard operations team will review the graphics. The review results include the following three categories: (1) project not yet considered, (2) supplement to existing graphic tutorials, (3) add new graphic tutorials.

The Cooperative Development module is built based on GitHub Pull Requests. All community members can participate in the development of the Bizard project through GitHub Pull Requests, such as contributing new document, optimizing the website, and project maintenance etc. The new graphic tutorials should be written in accordance with Bizard's formatting standards and should include the following: Graphic, Description, Example, Setup, Data Preparation, Visualization, Application, and Reference. The Bizard operations team will review the tutorial and provide suggestions for revisions.

The Automatic Deployment module is built based on GitHub Actions. To ensure efficient and reliable website development and completely eliminate the tediousness, delays, and risk of human error associated with manual website updates, Bizard has built a comprehensive automated workflow using GitHub Actions. This fully automates the entire process from code submission to website deployment, significantly improving the agility and stability of website iterations. The workflow is as follows: When a developer pushes new website content or code updates to a designated master branch (such as main or master) or initiates a pull request, GitHub Actions is automatically activated. It immediately executes a series of key tasks within a clean virtual environment provided by GitHub, strictly following a pre-defined configuration file (Workflow File), building the website and automatically deploying it to GitHub Pages.

## 4 Discussion

In this study, we developed Bizard, a community-driven visualization code storage and collaboration platform for the biomedical research. Compared to many existing visualization tools, Bizard's feature lies not only in the visualization capabilities it provides, but also in its integrated ecosystem of tools, data, tutorials, literature, and community. It aims to systematically address multiple core challenges faced by biomedical researchers in data visualization practice. Website visit statistics show that since its launch, Bizard platform website has received approximately 8,000 visits from more than 60 countries worldwide.

We firmly believe that leveraging Bizard's multi-platform resources will significantly enhance researchers' data visualization skills in the biomedical field. This will not only drive advancements in research methodologies but also improve the standards of data interpretation across the discipline. In turn, this progress will accelerate the development of precision medicine and personalized therapeutic interventions.

We encourage researchers to share their feedback and suggestions on both the GitHub and WeChat discussion communities. These valuable insights will play a crucial role in the platform's iterative optimization and functional expansion, helping to better meet the needs of the biomedical research community and expedite the translation from basic research to clinical applications [39].

In future developments, Bizard aims to expand its international network, attracting a diverse group of interdisciplinary researchers to collaborate and create more innovative and forward-thinking visualization solutions. Additionally, we will offer increasingly sophisticated and detailed code tutorials to address the growing complexity of data analysis needs in biomedical research and promote the integration of bioinformatics and clinical medicine [40]. We are confident that Bizard will become an evolutionary platform in advancing the standardization, normalization, and transparency of data visualization methodologies in the biomedical domain [41, 42]. This will contribute significantly to improving research quality and efficiency across the field and accelerate the clinical translation of biomedical discoveries for the benefit of human health [43].

## Author contributions

S.W., P.L., J.Z., and J.L. conceptualized the study, supervised the project, acquired funding, and reviewed and revised the manuscript. S.W. and P.L. additionally coordinated multi-institutional collaboration and data integration; P.L. led the computational framework design; S.W. oversaw the web repository architecture and long-term maintenance. H.Z. and K.L. managed project execution and operations, including task scheduling, figure and table generation, code repository construction, and website deployment, and drafted and revised the manuscript. The remaining authors contributed to task implementation, code optimization, and website validation. Authors are listed in order of overall contribution. In this study, co-first authorship (marked with †) is designated for individuals with early and sustained involvement who made core contributions (or equivalent) to at least five figures/tables. All authors reviewed and approved the final manuscript.

## Acknowledgments

We gratefully acknowledge the support provided by the Openbiox Bioinformatics Innovation Collaboration Group (https://github.com/openbiox/) and the Open Source Promotion Plan 2025 (https://summer-ospp.ac.cn/, Project ID: 2551d0158). This work was funded by the National Natural Science Foundation of China (Grants No. 82303953, No. 82300168), Hunan Provincial Natural Science Foundation of China (Grant No. 2025JJ40079), Shanghai Sailing Program (23YF1424400), The Innovative Research Team of High-level Local Universities in Shanghai, and Central South University Startup Funding.

## Ethics Statement

Not applicable (no human/animal subjects).

## Conflict of interest

The authors declare no competing interests.

## Data Availability Statement

The data that support the findings of this study are openly available at https://github.com/openbiox/Bizard and https://openbiox.github.io/Bizard/.

## Table and Figures

| Features | Bizard | FigureYa | R Graph Gallery |
|---|---|---|---|
| **Graphics** | | | |
| Number of graphic types | 65+ | 59 | 49 |
| Basic graphics | Yes | Yes | Yes |
| Omics graphics | Yes | Yes | No |
| Clinical graphics | Yes | Yes | No |
| **Website** | | | |
| Language | Chinese+English | Chinese+English | English |
| Release year | 2025 | 2025 | 2019 |
| Graphics hierarchical menu | Yes | No | No |
| Real world data | Yes | Yes | No |
| Data cloud Link | Yes | No | No |
| **Community** | | | |
| Community-driven project | Yes | Yes | Yes |
| Comment system | Yes | No | No |
| Automated deployment | Yes | Yes | Yes |
| Contribution score mechanism | Yes | No | No |

**Table 1. Comparison between Bizard and Similar Data Visualization Platforms.**

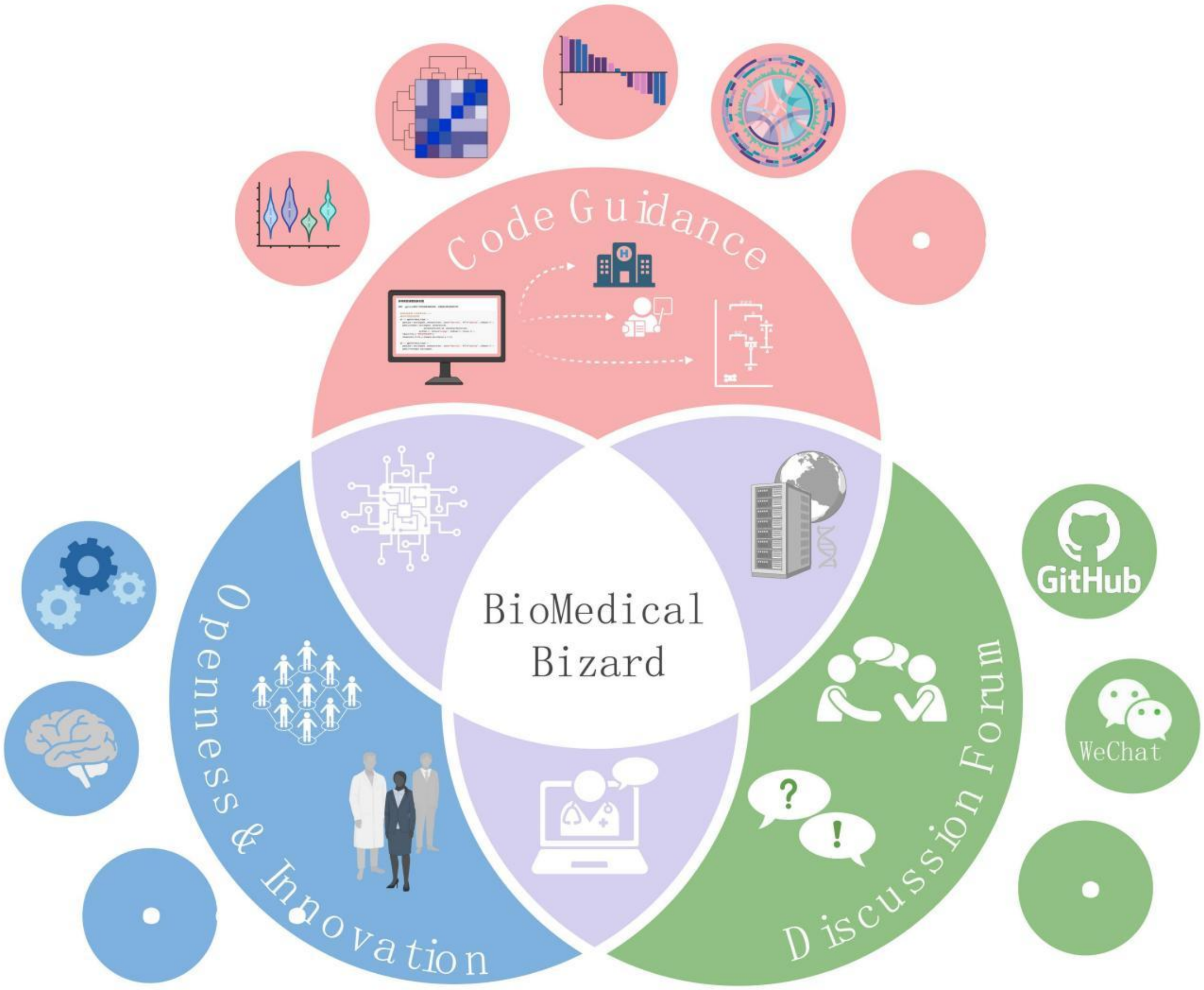

**Figure 1. The Overview of Bizard Project.**

The overview of Bizard, a comprehensive, community-driven platform that integrates code guidance, open innovation, and discussion forums for biomedical data visualization.

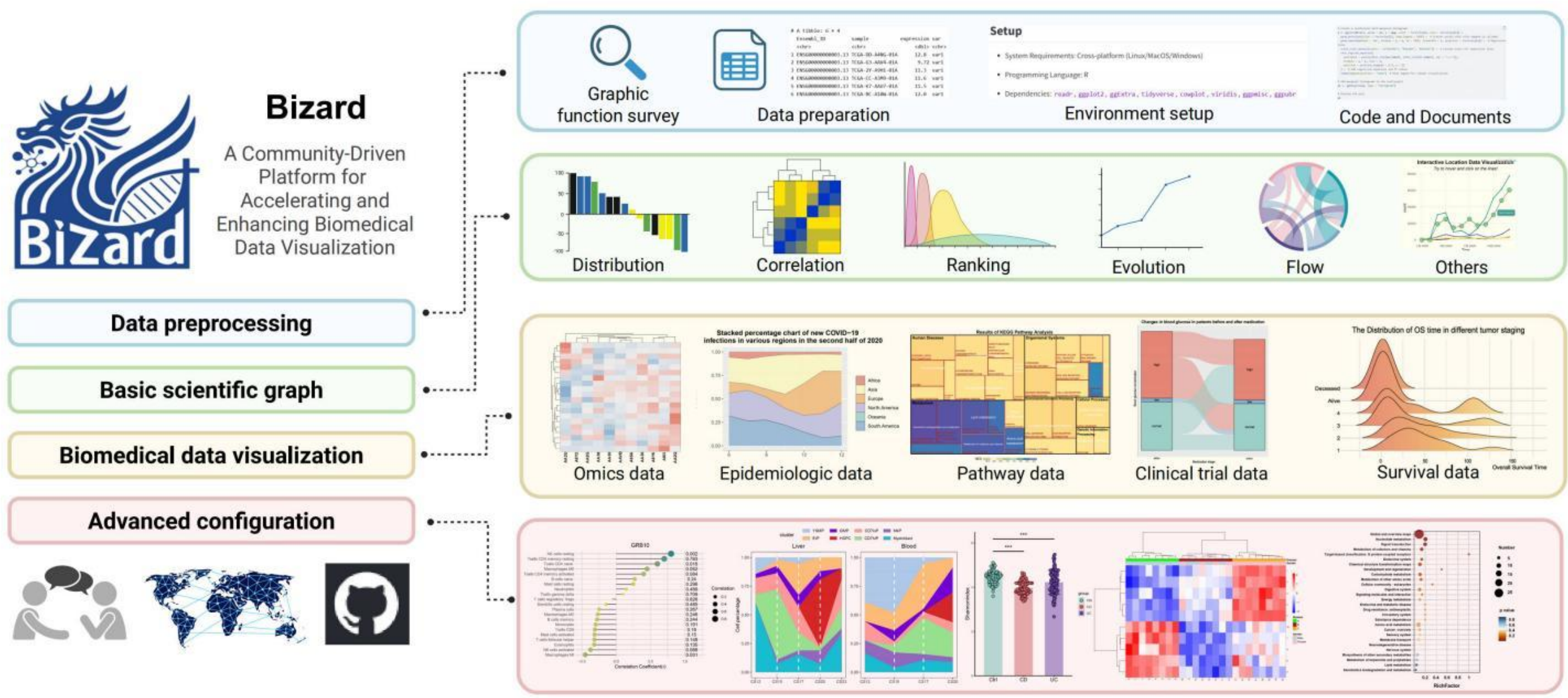

**Figure 2. The Workflow of Bizard's Method and Results.**

The left panel displays the core workflow of Bizard, while the right panel shows detailed content for different stages. Data preprocessing: Includes steps such as graphic function survey, data preparation, environment setup, and writing code and documents. Basic scientific graph: Includes basic plots such as distribution plots, correlation plots, ranking plots, evolution plots, flow plots, and others. Biomedical data visualization: Includes charts related to biomedical data, such as omics data, epidemiologic data, pathway data, clinical trial data, survival data, and others. Advanced configuration: Includes more advanced plots, such as charts available in Hiplot.

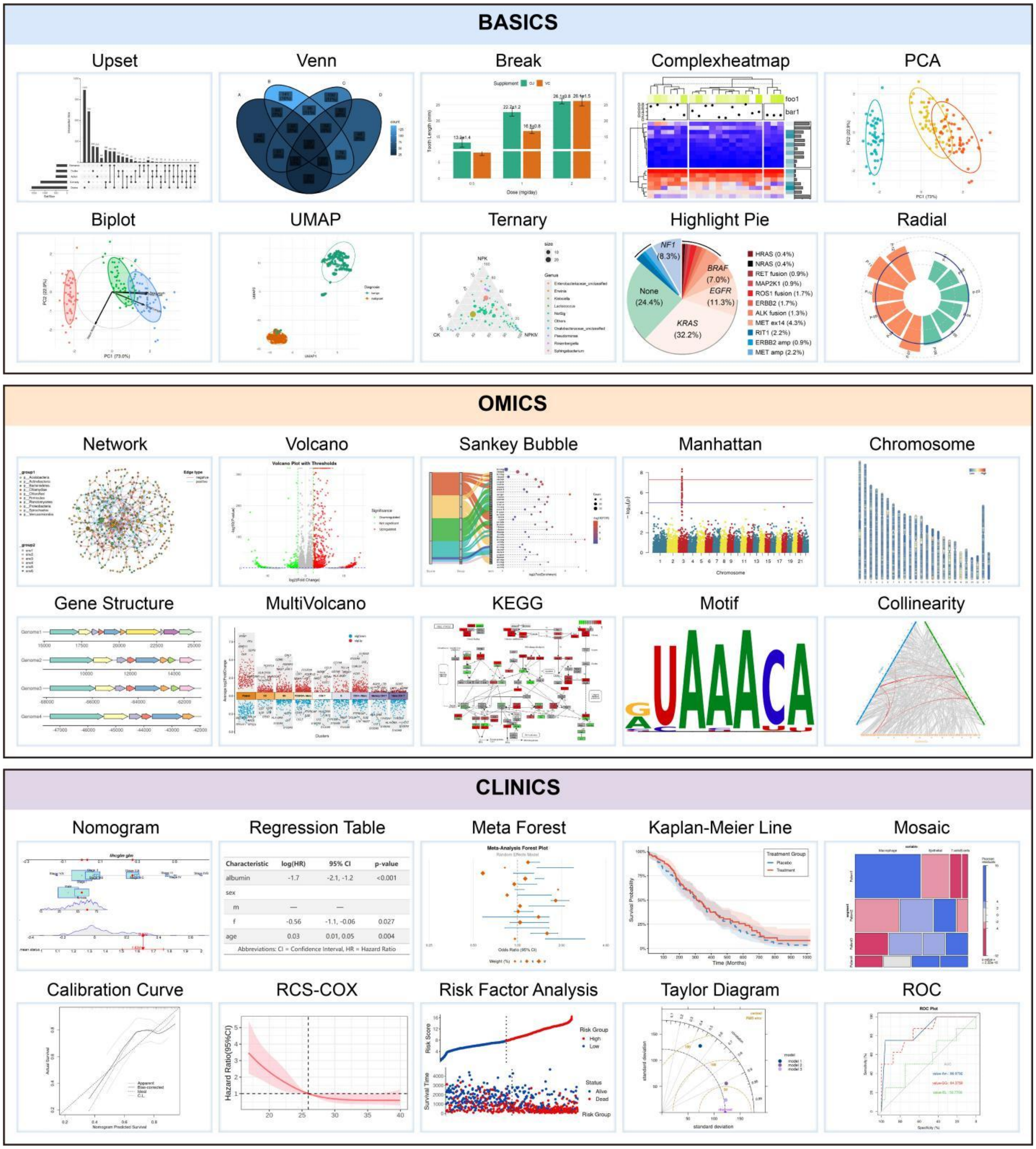

**Figure 3. Examples and Classification of Graphics in Bizard.**

This diagram shows the graphic classifications in Bizard, 10 example graphics are displayed for each classification. Basic category includes Upset plot, Venn plot, Pie chart and other basic graphics. Omics category includes Volcano plot, Chromosome plot, KEGG plot and other omics graphics for genome and transcriptome analysis. Clinics category includes Nomogram, Meta Forest, Kaplan-Meier Line and other clinics graphics for clinical medical analysis.

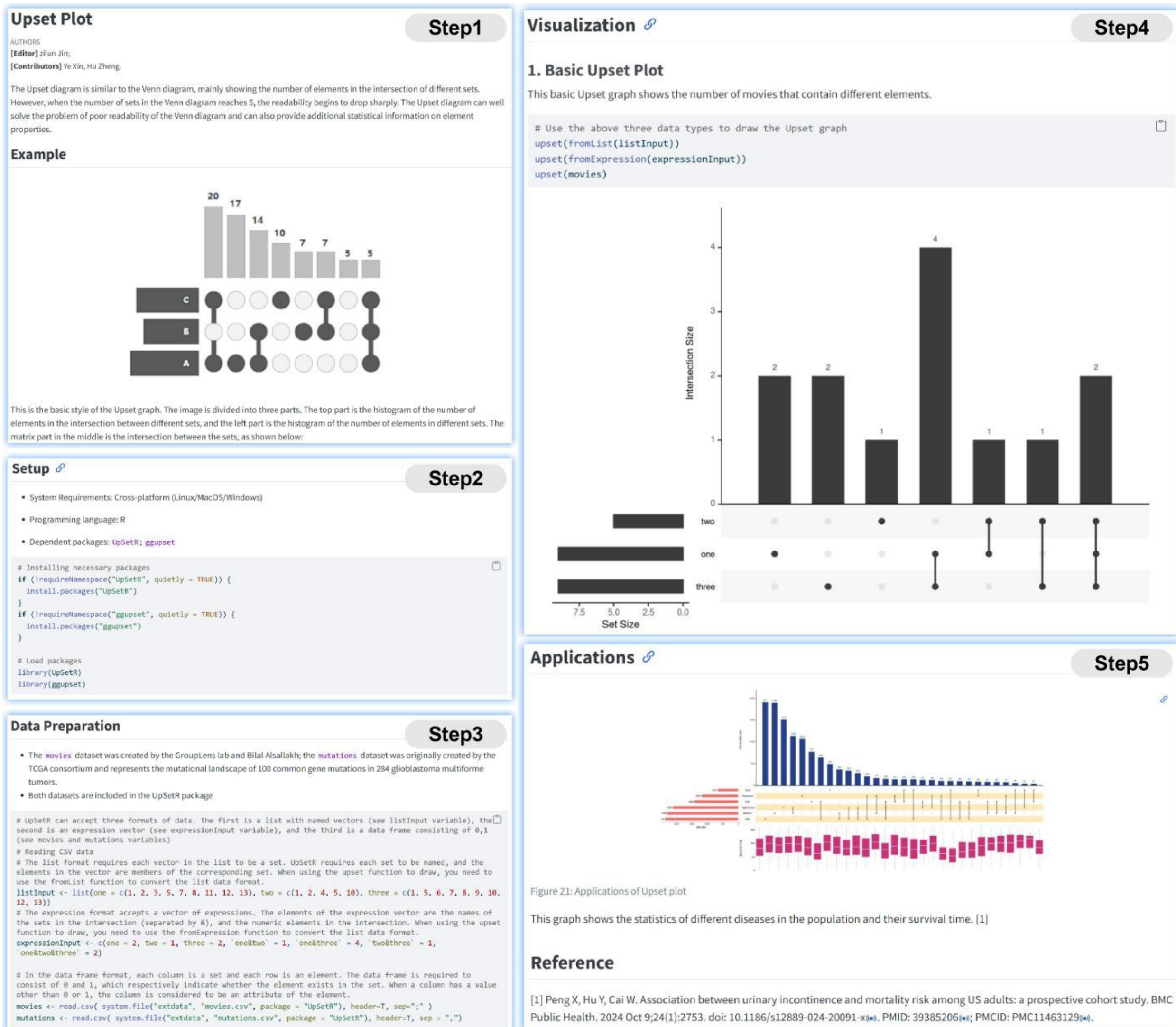

**Figure 4. Usage Workflow of Upset Plot Plugin.**

This diagram shows the five steps for drawing an Upset plot in Bizard. Step1: An overall introduction and example usage of Upset plot. Step2: Install and load the R packages used in this tutorial. Step3: Load the prepared online test dataset. Step4: Draw Upset plot using R code. Step5: Applications of Upset plot in published literature.

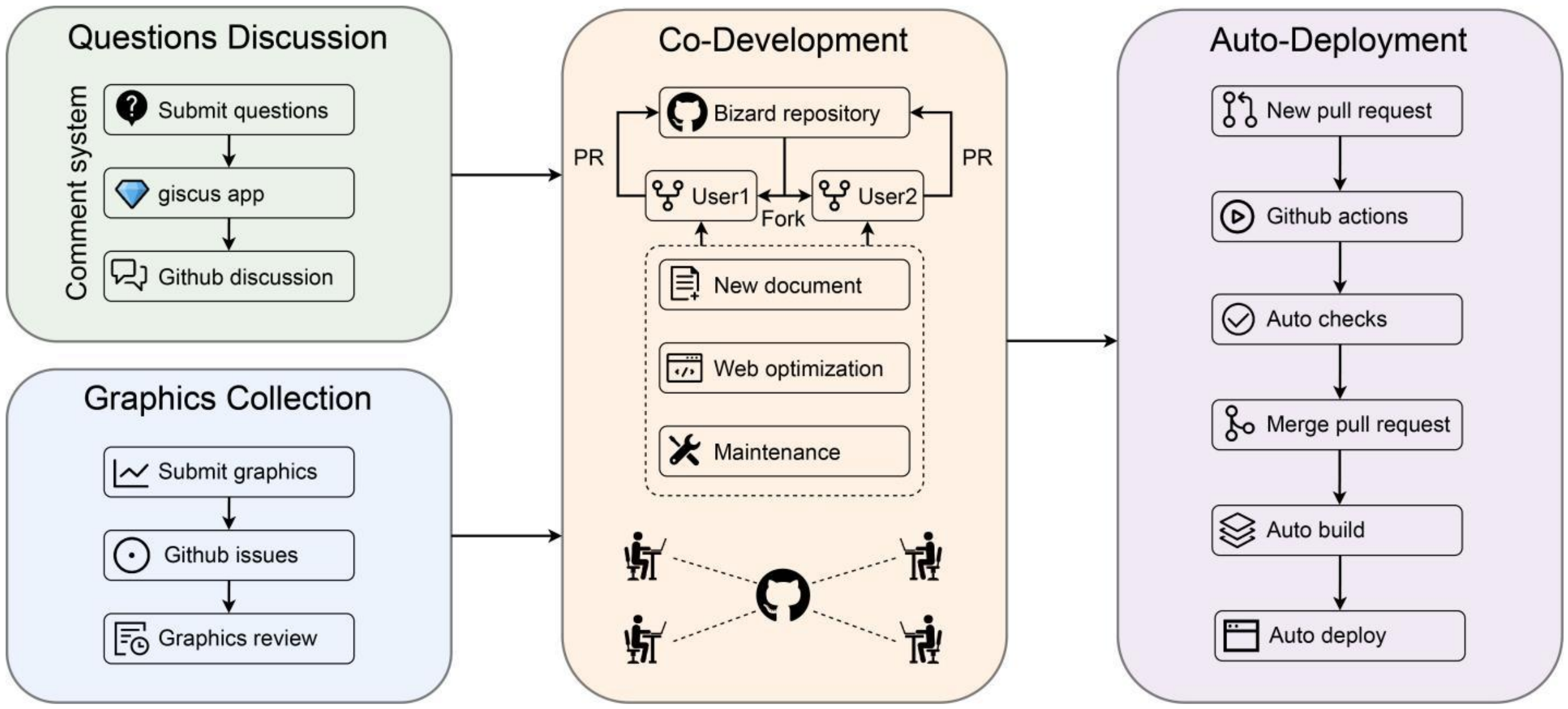

**Figure 5. Bizard Community Collaboration Framework.**

This diagram illustrates the four modules of Bizard community collaboration framework: questions discussion (green), graphics collection (blue), cooperative development (orange), and automatic deployment (purple). The questions discussion module based on github discussion is used for community users to discuss problems; the graphics collection module based on github issues is used for collecting new graphics from community users; the cooperative development module based on github pull request is used for collaborative development among community users; and the automatic deployment module based on github actions is used for automatical checks and deployment of project websites.